\documentclass[journal=jacsat,manuscript=article]{achemso}

\usepackage{chemformula} 
\usepackage[T1]{fontenc} 
\usepackage{tabularx}

\author{Sreejani Karmakar}
\affiliation{Department of Physics, Indian Institute of Science Education and Research (IISER) Tirupati, Tirupati - 517507, Andhra Pradesh, India.}
\email{sreejanikarmakar@students.iisertirupati.ac.in}
\author{Sudipta Dutta}
\affiliation{Department of Physics and Center for Atomic, Molecular and Optical Sciences \& Technologies, 
Indian Institute of Science Education and Research (IISER) Tirupati, Tirupati - 517507, Andhra Pradesh, India.}
\email{sdutta@iisertirupati.ac.in}
\phone{+91 877 2500 434}

\title[An \textsf{achemso} demo]
  {Strain Tunable Photocatalytic Ability of $BC_{6}N$ Monolayer}

\begin{document}


\begin{abstract}
We explore the photocatalytic ability of honeycomb lattice of borocarbonitride, $BC_{6}N$, investigating its opto-electronic properties and band-edge alignment based on hybrid density functional theory. We observe that along with pronounced visible absorbance, $BC_{6}N$ exhibits a good reducing ability. It is capable of depositing heavy metal ions, hydrogen fuel generation and carbon dioxide sequestration through photoelectrocatalysis. To broaden its applicability we further tune its redox ability using biaxial strain. We observe even a slight expansion makes this material capable of performing simultaneous oxidation and reduction, which is essential for waste management and hydrogen generation through spontaneous water splitting. In this study, we propose $BC_{6}N$ to be a promising potential candidate as metal free 2D photocatalyst and show ways to tune its applicability to address multiple environmental crises.
\end{abstract}


\section{Introduction}

Photocatalysis is an important pillar to build a strategic transition towards a green and sustainable future \cite{1,2,3}. Photo-redox reactions are experimentally proven to be used for hydrogen fuel production, waste water management, reduction of heavy metal ions, degradation of organic pollutants , carbon dioxide reduction and many more, without any detrimental effect on nature  \cite{1,2,3,4,5,6,7}. After the experimental realization of solar water splitting in 1972 \cite{4}, numerous revolutionary theoretical and experimental studies have been conducted in this field but the achieved photo-conversion efficiency is still far from commercial viability \cite{8,9,10,11,12,13}. So the search for mechanically strong, stable semiconductor with suitable opto-electronic property and proper band alignment for the required redox reaction is still on. Our work aims to contribute to this global search for an appropriate photocatalyst.

Nanostructured semiconductors are established to provide quite a few advantages over bulk semiconductors for photocatalysis. For example, high surface to volume ratio of nanostructured semiconductors results in an increased number of active sites for catalytic reaction and increased area for photon harvest. Moreover due to high carrier mobility and reduced migration path length the recombination of the photogenerated charge carriers are suppressed significantly in reduced dimension. Also, its easily tunable optoelectronic property allows one to tune the redox ability efficiently \cite{14,15,Dutta}. Among other reduced dimensional (1D or 0D) materials, 2D materials give extra advantage as it is comparatively easier to handle and reuse. More importantly standard sophisticated techniques are already established to produce a broad range of 2D materials \cite{Andrew}. Consequently, the advancement of 2D materials has significantly strengthened photocatalytic technology \cite{16,17,18}. 

Here, we explore the photocatalyitic ability of one such theoretically predicted 2D material called graphene like borocarbonitride. $BC_{6}N$. The most stable configuration of it is shown in Fig.1(a) where each carbon hexagon is surrounded by another hexagon made up of alternating boron and nitrogen \cite{19}. It can be thought of an intermediate of graphene and hexagonal boron nitride (hBN). hBN has alternating boron and nitrogen in its honeycomb hexagon and exhibits large band gap (5.9 eV) \cite{20}, which is far from semimetallic nature of graphene, consisting of only carbon atoms. Both theoretical and experimental studies show that a hybrid structure of hBN and graphene can lead to unique and versatile materials with properties intermediate of these two and consequently have the potential to overcome the drawbacks of these individual materials in application purpose \cite{21,22,Dutta1,23,24}. This fact is confirmed by the reported properties of $BC_{6}N$. It has a moderate direct band gap (1.83 eV) \cite{19,25} which is highly beneficial for optoelectronic application. Due to the fact that it is made up of boron, carbon, nitrogen, three neighboring elements in periodic table with comparable atomic sizes and a tendency to make strong covalent bond among each other, $BC_{6}N$ exhibits high mechanical strength \cite{25}. Moreover, since $BC_{6}N$ consists of only nonmetal p-block element, it is advantageous over metal-based semiconductors which suffer from the drawback of poor stability, high cost and high toxicity \cite{26}. $BC_{6}N$ is also predicted to have high carrier mobility, indicating low charge recombination in photocatalysis  \cite{19,27}. These fascinating predicted properties have drawn significant experimental attention to this kind of materials. Recently Matsui et al experimentally synthesized BN codoped nanographene with a similar atomic configuration \cite{28}. Owing to these favorable properties along with the recent experimental advances, we decide to conduct a quantitative exploration of the ability of $BC_{6}N$ as a potential candidate for metal-free 2D photocatalyst.

\section{Computational details}

The density functional theory (DFT) based calculations are done using Vienna Ab Initio Simulation Package (VASP) \cite{vasp1,vasp2}. First the structural relaxation is performed along with the optimization of lattice vectors, using projector augmented-wave (PAW) potentials and Perdew-Burke-Ernzerhof (PBE) exchange and correlation functional under generalized gradient approximation (GGA) \cite{vasp3}. But since it underestimates band gaps, Heyd-Scuseria-Ernzerhof (HSE06) hybrid functional with 25\% exact exchange is used for electronic and optical calculations \cite{Heyd}. The calculations are performed with plane wave basis set with cutoff energy of $520$ eV and 20\AA vacuum is created in the non-periodic direction to avoid the interactions between two adjacent unit cells. A $5\times5\times1$ Monkhorst-Pack k-point grid is used for Brillouin zone sampling. Electronic self consistency cut-off of $10^{-6}$ eV is considered and force tolerance on each atom has been converged upto $0.01$ eV/\AA. Using VASP code the imaginary part of the dielectric tensor is calculated by summing over the empty states using the equation,

 \begin{equation}
    \varepsilon_{\alpha \beta}^{(I)}(\omega) = \frac{4\pi^{2}e^{2}}{\Omega} \lim_{q\to 0} \frac{1}{q^2} \sum_{c,v,\textbf{k}} 2\omega_{\textbf{k}} \delta(\epsilon_{c\textbf{k}}-\epsilon_{v\textbf{k}}-\omega) <u_{c\textbf{k}+e_{\alpha q}}|u_{v\textbf{k}}><u_{c\textbf{k}+e_{\beta q}}|u_{v\textbf{k}}>^{*}
 \end{equation}

\noindent where $c$ and $v$ refer to the conduction and valence band, respectively and $u_{c\textbf{k}}$ is the cell periodic part of the orbitals at the wave vector \textbf{k}. The real part is obtained using Kramers-Kronig relation,

\begin{equation}
    \varepsilon_{\alpha \beta}^{(R)}(\omega) = 1 + \frac{2}{\pi} P \int_{0}^{\infty}\frac{\varepsilon_{\alpha \beta}^{(I)}(\omega^\prime)\omega^\prime}{{\omega^\prime}^2 - \omega^2 + i\eta}d\omega'
\end{equation}

\noindent where $P$ indicates the principle value. VASPKIT, a post-processing program for VASP code \cite{29} is used to calculate absorption coefficient from the information of real and imaginary part of the dielectric constant using the following formula,

\begin{equation}
    \alpha_{\alpha\beta}(\omega) = \frac{\sqrt{2}\omega}{c}[\sqrt{(\varepsilon_{\alpha\beta}^{(I)})^{2}+(\varepsilon_{\alpha\beta}^{(R)})^{2}} - \varepsilon_{\alpha\beta}^{(I)}]^{\frac{1}{2}}
\end{equation}

The band edge alignment with respect to hydrogen reduction and water oxidation level is calculated from vacuum level reference. From the information of local potential as obtained using VASP, vacuum level energy is calculated using VASPKIT. The work function $(\phi)$ is obtained by subtracting the Fermi energy $(E_{f})$ from vacuum energy $E_{Vacuum}$. Position of the Fermi energy with respect to vacuum level is given by $-\phi$. With the knowledge of Fermi energy level and the band positions with respect to Fermi energy as obtained in band structure, band edge alignment is calculated.

\section{Results and discussion}

The relaxed borocarbonitride ($BC_{6}N$) exhibits a completely planer honeycomb structure (shown in Fig.1(a)). The lattice vectors $\vec{a}$ and $\vec{b}$ are 4.97\AA with an angle of $60^{\circ}$ between them. The bond angles are nearly equal to $120^{\circ}$. All the relaxed structure parameters match with earlier reported literature \cite{19}. Study of electronic dispersion relation reveals that $BC_{6}N$ is a direct band gap semiconductor with appearance of both conduction band minima (CBM) and valence band maxima (VBM) at high-symmetric K point (shown in Fig.1(b)). The value of the band gap as obtained with GGA-PBE functional is 1.27 eV and the same with HSE06 hybrid functional increases to 1.83 eV, which are in well agreement with previous literature \cite{19}. For this class of materials, HSE06 is reported to give a more accurate result \cite{Muscat} and hence for the rest of the calculations only HSE06 functional is used.

\begin{figure}[!ht]
    \centering
    \includegraphics[width=0.8\linewidth]{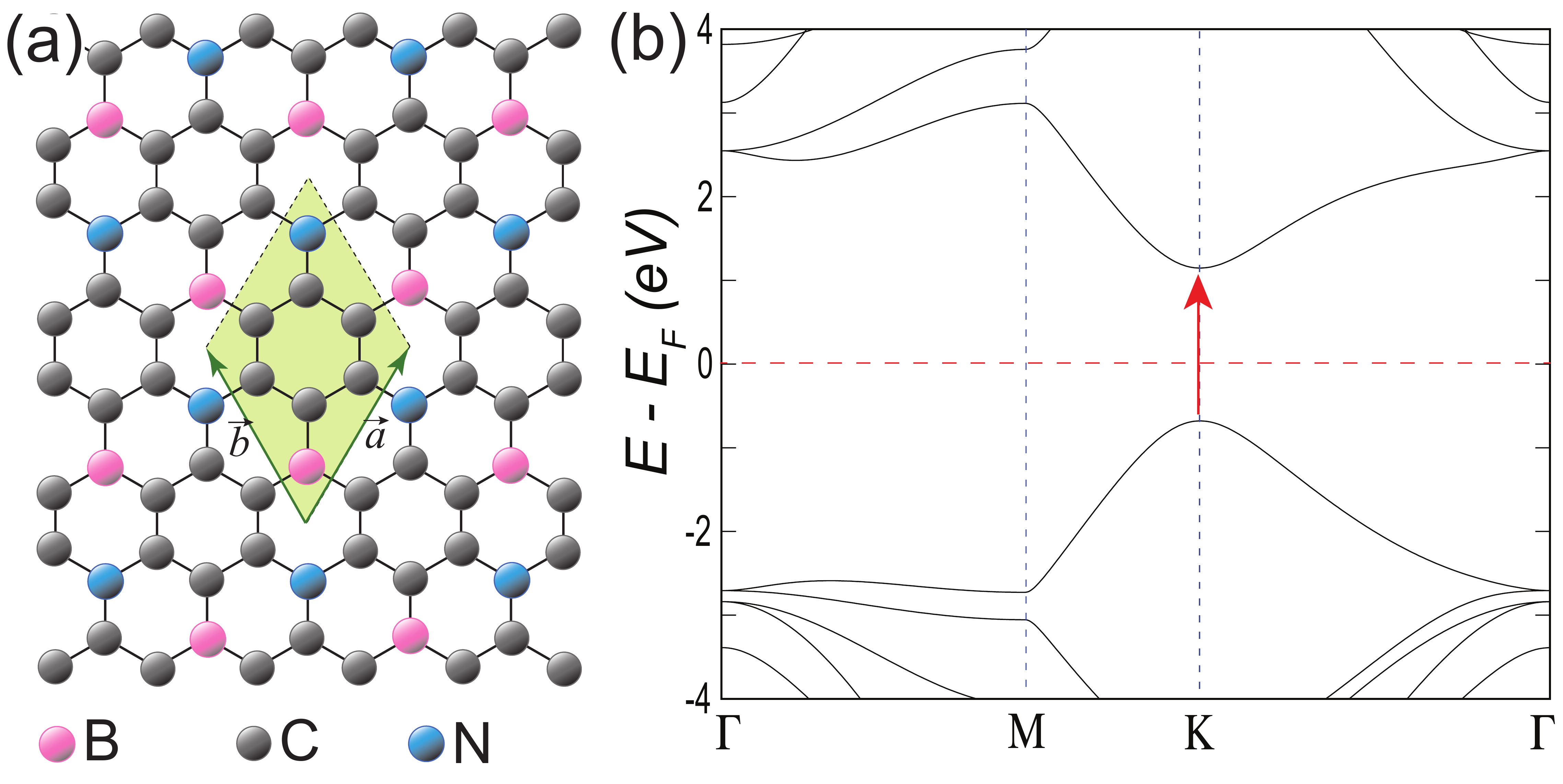}
    \caption{(a) Structure of monolayer $BC_{6}N$ system. The shaded rhombus indicates the unit cell with lattice vectors, $\vec{a}$ and $\vec{b}$. (b) Electronic energy dispersion of monolayer $BC_{6}N$ as obtained with HSE06 functional. The vertical (horizontal) lines indicate the high-symmetric points (Fermi energy). The arrow at high-symmetric point K shows the direct band gap of 1.83 eV.}
\label{fig:my_label}
\end{figure}

The absorption spectra (shown in Fig.2(a)) is explored to have an accurate estimate of energy of the harvested photon. We observe the absorption edge at 1.83 eV and hence the optical band gap is same as electronic band gap. This implies even the minimum energy of the harvested photon is greater than the net free energy change in photocatalytic water splitting (1.23 eV)\cite{Walter} and the harvested photons will be capable of driving the catalytic reaction. Also, there are pronounced peaks in the visible range. As the solar spectra peaks at the visible range, this result ensures high photon harvest under solar irradiation which is very crucial to maximize efficiency.

\begin{figure}[!ht]
    \centering
    \includegraphics[width=0.8\linewidth]{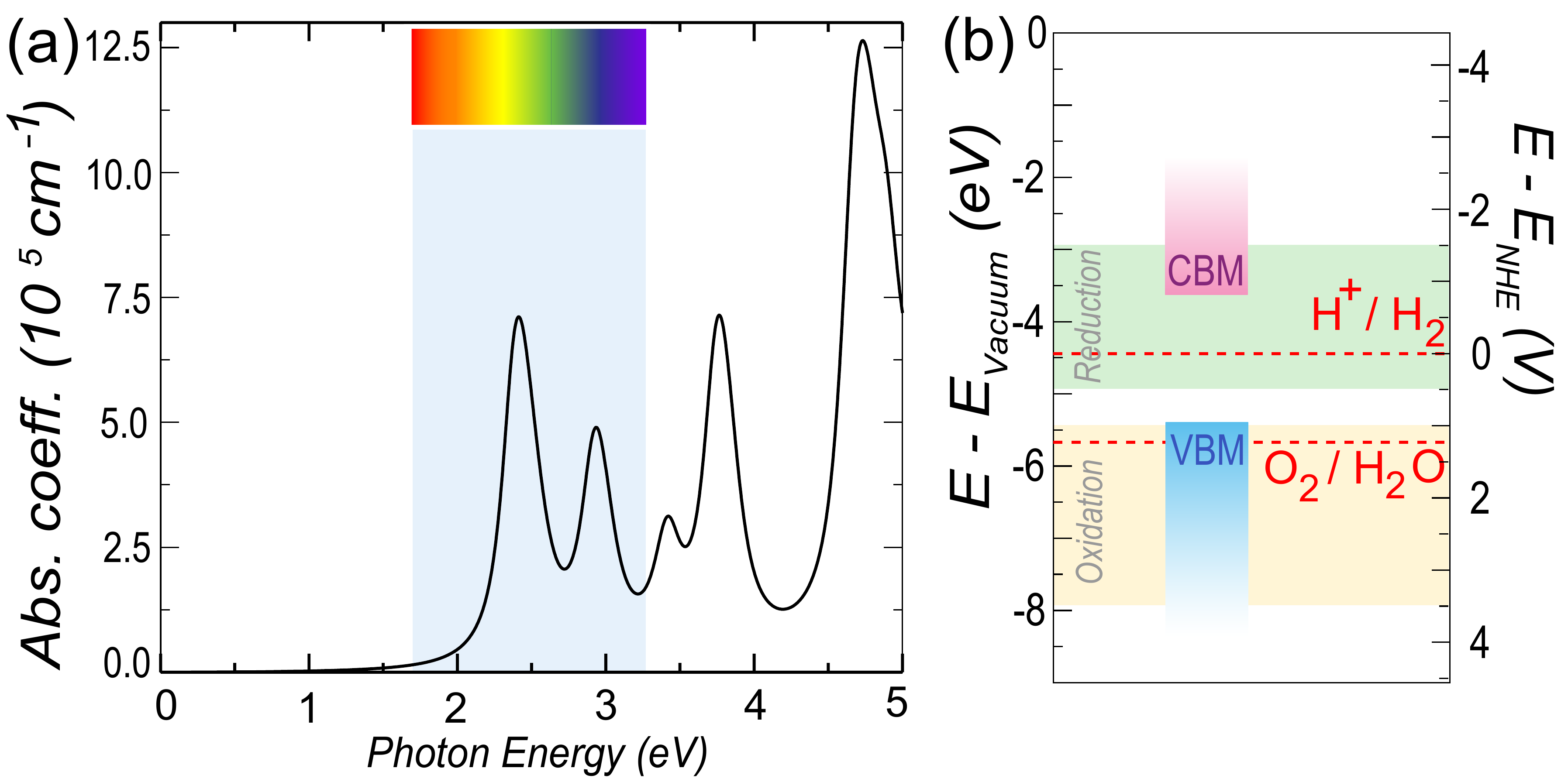}
    \caption{(a) Absorption spectra of monolayer $BC_{6}N$. The shaded region schematically depicts the visible spectrum. (b) The valence band maxima (VBM) and conduction band minima (CBM), i.e., the band edge alignment of monolayer $BC_{6}N$, as obtained with HSE06 exchange correlation functional. The vertical left axis represents the energy scaled with respect to vacuum energy ($E_{Vacuum}$), and vertical right axis indicates the potential value with respect to Normal Hydrogen Electrode potential ($E_{NHE}$). The potentials for hydrogen reduction (H$^{+}$/H$_{2}$) and water oxidation (O$_{2}$/H$_{2}$O) are depicted by horizontal dashed lines. The range for optimal reduction and oxidation energetics are shown by the shaded regions.}
 \label{fig:my_label}
\end{figure}

But for exact quantitative estimation about the photocatalytic ability of any material, the information regarding its band edge positions with respect to the redox potentials is very crucial. To have a spontaneous electron transfer during reduction or oxidation, the chemical potential of the conduction electrons must be lower than the potential required for reduction and the same for valence holes must be higher than the oxidation potential \cite{Walter}. The offset determines the thermodynamic drive for corresponding reduction or oxidation. The band edge positions of $BC_{6}N$ are shown in Fig.2(b). The relative position of the band edge with respect to Normal Hydrogen Electrode (NHE) is obtained by using the value of standard redox potential with respect to vacuum, which are -4.44 V and -5.67 V for hydrogen reduction and water oxidation respectively \cite{30,31}. For optimal reducing and/or oxidizing performance, the material should have its conduction electrons at a chemical potential of +0.5V to -1.5V (vs $E_{NHE}$) and/or valence band holes with a chemical potential of +1.0V to +3.5V (vs $E_{NHE}$), respectively \cite{32}. The shaded regions in Fig.2(b) indicate the above mentioned optimal ranges. We find that the position of the conduction band is at -3.60 eV with respect to vacuum (+0.84V vs $E_{NHE}$), providing a reducing offset of 0.84 eV. The valence band is at -5.43 eV with respect to vacuum (0.99V vs $E_{NHE}$) and hence is slightly above the optimal range. In the presence of an appropriate hole scavenger, the reducing ability of $BC_{6}N$ can be exploited to deposit heavy metal ions from contaminated water in order to purify it. It can also be used as a photocathode for hydrogen fuel generation or carbon dioxide sequestration through photoelectrocatalysis. 

\begin{figure}[!ht]
    \centering
    \includegraphics[width=0.8\linewidth]{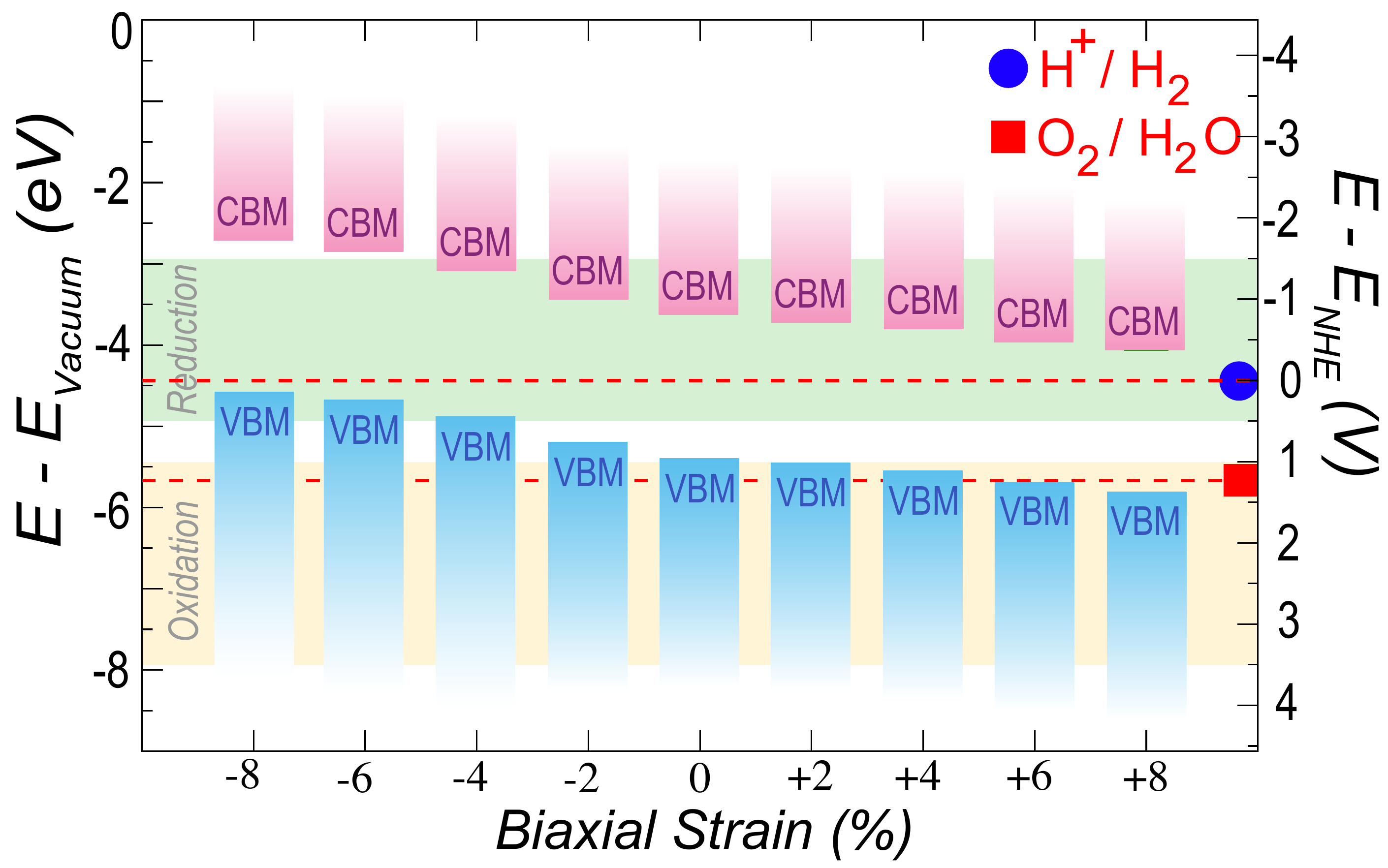}
    \caption{Variation of band edge positions of monolayer $BC_{6}N$ with applied biaxial strain. The potentials for hydrogen reduction (H$^{+}$/H$_{2}$) and water oxidation (O$_{2}$/H$_{2}$O) are depicted by horizontal dashed lines with a filled circle and a square, respectively. The range for optimal reduction and oxidation energetics are shown by the shaded regions.}
 \label{fig:my_label}
\end{figure}

In spite of good reducing performance, the absence of oxidation offset limits applicability of $BC_{6}N$. To broaden its applicability we further tune the band alignment and hence redox ability by applying biaxial strain. Experimentally also this can be easily achieved by applying external load or through lattice mismatch \cite{33,34,35}. Here the biaxial strain is quantified through the parameter $\delta = (a_{s}-a_{0})/a_{0}$, where $a_{0}$ is the relaxed lattice constant and $a_{s}$ is the same after applying strain. The effect of this biaxial strain on band edges of the material is shown in Fig.3 and the corresponding data are given in Table.1. This study shows that the value of the band gap does not change significantly with strain. Though it increases with compression (negative strain) and decreases with expansion (positive strain), it continues to be in the visible range throughout. Whereas the work function is quite sensitive to the applied strain. It decreases with compression and increases with expansion. The trend of variation in the work function is opposite and more rapid as compared to the change in band gap. As a consequence, in spite of the decreasing trend of band gap, due to increasing work function, the relative position of the Fermi energy goes down with expansion and the material manages to meet the criteria on both the conduction electrons and valence holes, to be called a good reducing material and good oxidant simultaneously. 2\% expansion onward the oxidation offset appears and it increases with an increase in expansion. This oxidizing ability of holes can be utilized to degrade a wide variety of organic pollutants \cite{32}. After +6\% the valence band maxima just crosses the water oxidation and at around +8\% it grows prominent offset for spontaneous water oxidation. Under this expansion range, $BC_{6}N$ can oxidize water into $H^{+}$ and $O_{2}$ and reduce $H^{+}$ to $H_{2}$ simultaneously, producing $H_{2}$ fuel through spontaneous water splitting. The conduction electrons can also be used to reduce $O_{2}$ to ${O_{2}}^{-}$ which has moderate disinfectant property and can be used to degrade bacteria \cite{Yuan}. With compression, the reducing offset increases however the applicability is similar to that of unstrained $BC_{6}N$. These results show that by applying biaxial strain, $BC_{6}N$ can be utilized for a wide range of application. Moreover this strain tunable redox ability can be used to suppress multiple side reactions, giving one a good control over reaction pathways, allowing to use the material for any particular required application with increased efficiency.

\begin{center}
    
\begin{table*}[h]
\small
\caption{Effect of biaxial strain on the band gap (using both PBE and HSE06 exchange correlation functionals), work function (using HSE06) and band edge positions (using HSE06), i.e. the position of valence band edge (VBE) and conduction band edge (CBE) with respect to (w.r.t) the vacuum energy and with respect to the Normal Hydrogen Electrode potential of monolayer $BC_{6}N$ system.}
\label{tab:table4}
  \begin{tabularx}
{1\textwidth} {|> {\raggedright\arraybackslash}X |>  {\raggedright\arraybackslash}X |> {\raggedright\arraybackslash}X |> {\raggedright\arraybackslash}X |> {\raggedright\arraybackslash}X |> {\raggedright\arraybackslash}X |> {\raggedright\arraybackslash}X |> {\raggedright\arraybackslash}X |}
\hline
Strain (\%) & PBE gap (eV) & HSE06 gap (eV) & Work function (eV) & CBE w.r.t vacuum (eV) & VBE w.r.t vacuum (eV) & CBE w.r.t $E_{NHE}$ (V) & VBE w.r.t $E_{NHE}$ (V) \\
\hline
-8 & 1.38 & 1.92 & 4.16 & -2.68 & -4.60 & -1.76 & +0.16 \\
\hline
-6 & 1.36 & 1.89 & 4.27 & -2.82 & -4.71 & -1.62  & +0.27 \\
\hline
-4 & 1.32 & 1.87 & 4.41 & -3.05 & -4.92 & -1.39  & +0.48 \\
\hline
-2 & 1.30 & 1.83 & 4.62 & -3.40 & -5.23 & -1.04  & +0.79 \\
\hline
 0 & 1.27 & 1.83 & 4.73 & -3.60 & -5.43 & -0.84  & +0.99 \\
\hline
+2 & 1.25 & 1.80 & 4.79 & -3.69 & -5.49 & -0.75  & +1.05 \\
\hline
+4 & 1.23 & 1.80 & 4.89 & -3.77 & -5.57 & -0.67  & +1.13 \\
\hline
+6 & 1.21 & 1.79 & 4.95 & -3.93 & -5.72 & -0.51  & +1.28 \\
\hline
+8 & 1.20 & 1.79 & 5.10 & -4.04 & -5.83 & -0.40  & +1.39 \\
\hline
\end{tabularx}
\end{table*}
\end{center}

\section{Conclusions}

Our results show that the $BC_{6}N$ is a good reducing material and can serve as a photocathode in photoelectrocatalysis. Compression increases its reducing ability whereas with expansion additional oxidizing ability appears. Depending on the extent of biaxial strain, $BC_{6}N$ can address multiple environmental issues. It can be used for water purification in various different ways, like reductive deposition of heavy metal ions, oxidative degradation of organic pollutants, disintegration of bacteria etc. It can contribute to air pollution control through carbon dioxide sequestration. Beyond 6\% tensile strain, it can also help in the remediation of global energy crisis by producing hydrogen fuel with zero carbon footprint. Therefore, we claim $BC_{6}N$ to be a potential candidate as metal free 2D photocatalyst with pronounced visible absorption and its photocatalytic ability is worthy of experimental exploration for remediation of environmental crises.

\begin{acknowledgement}
SK and SD thank IISER Tirupati for Intramural Funding and Science and Engineering Research Board 
(SERB), Dept. of Science and Technology (DST), Govt. of India for Early Career Research (ECR) award grant 
(ECR/2016/000283). The support and the resources provided by ‘PARAM Brahma Facility’ 
under the National Supercomputing Mission, Government of India at the Indian Institute of Science Education 
and Research (IISER) Pune are gratefully acknowledged.
\end{acknowledgement}

\end{document}